\documentclass{iagproc}

\title{Empirical Earth rotation model: a consistent way 
       to evaluate Earth orientation parameters.}
\author{L. Petrov \\ NVI, Inc./NASA GSFC, Code 698, Greenbelt, 
         20771 MD, USA }

\begin{document}

\maketitle

\begin{abstract}
  It is customary to perform analysis of the Earth's rotation in two 
steps: first, to present results of estimation of the Earth orientation 
parameters in the form of time series based on a simplified model of 
variations of the Earth's rotation for a short period of time, and then 
to process this time series of adjustments by applying smoothing, re-sampling 
and other numerical algorithms. Although this approach saves computational 
time, it suffers from self-inconsistency: total Earth orientation 
parameters depend on a subjective choice of the apriori Earth orientation 
model, cross-correlations between points of time series are lost, and results 
of an operational analysis per se have a limited use for end users. 
An alternative approach of direct estimation of the coefficients of expansion
of Euler angle perturbations into basis functions is developed. These 
coefficients describe the Earth's rotation over entire period of observations
and are evaluated simultaneously with station positions, source coordinates 
and other parameters in a single LSQ solution. In the framework of this 
approach considerably larger errors in apriori EOP model are tolerated. This
approach gives a significant conceptual simplification of representation 
of the Earth's rotation. 
\end{abstract}

\begin{keywords}
  astrometry, reference systems, Earth rotation, VLBI
\end{keywords}

\section{Introduction}

  The complexity of the traditional formalism for describing the Earth's 
rotation is frustrating. Even in the community of professional astronomers 
a complete understanding of the procedure of applying reduction for the 
Earth's rotation in all details is not common, especially of the last 
IAU~2000, IAU~2006 recommendations \citep{r:iers2003}. However, reduction 
for the Earth's rotation is not only a subject of academic interest, 
but is the area for various important applications. The complexity of the 
formalism increases the probability of misunderstanding. Misunderstanding 
significantly increases the risk of an error in a software which implements 
reductions. In the worst case scenario, for instance, a glitch in 
a navigation equipment of a passenger aircraft controlled by such 
a program may result in a wreck and loss of life. That is why it is vitally 
important to simplify description of the Earth's rotation for end users. 

   An alternative approach for describing the Earth rotation was proposed
by \citet{r:erm}. It was demonstrated that instead of using a very complex 
mathematical model for the a~priori rotation matrix and time series of the 
Earth orientation parameters which correct that a~priori rotation matrix, 
it is feasible to represent the Earth's rotation in a form of a sum of
a very simple, coarse a~priori mathematical model and a set of coefficients
of expansion of the perturbational rotation into the Fourier and B-spline 
bases which are estimated from observations. 

   In this paper the implications of the alternative approach with respect 
to the traditional approach, its advantages and limitations, are discussed. 

\section{Definition of the Earth's rotation}

   Space geodesy techniques, such as GPS, SLR, DORIS and VLBI allow 
to measure the time intervals or the differences of time intervals of 
electromagnetic wave propagation from observed bodies to observing stations.
Solutions of differential equations of wave propagation ties position vectors
of observed bodies and their time derivatives with position vector of 
observing bodies and their time derivatives. Therefore, a relative motion 
of observing stations with respect to observed bodies can be determined 
from analysis of observations. Since the observing stations are fixed 
with respect to the Earth's crust, it is convenient to express their 
positions in a terrestrial coordinate system in which a motion of stations 
would be small, an order of magnitude of Earth's crust deformation, 
i.e. $10^{-8}$. A time-independent reference position vector in such 
a coordinate system, $\vec{r}_{{}_T}$ is determined via the following 
linear transformation:
{\small
\begin{eqnarray}
   \vec{r}(t)_{{}_C} = \widehat{\mathstrut\cal M}_a(t) \, \vec{r}_{{}_T} +
                    \vec{q}_e(t) \times \vec{r}_{{}_T} + \vec{d}_a(t){{}_T} + 
                    \vec{d}_e(t){{}_T} \hspace{-0.5em}
\label{e:01}\end{eqnarray}
}
   here $\vec{r}(t)_{{}_C}$ is the reference position vector in celestial 
coordinate system, $\vec{d}_a(t){{}_T}$ and $\vec{d}_e(t){{}_T}$ are the
time-dependent a~priori and empirical station displacement vectors, 
$\widehat{\mathstrut\cal M}_a(t)$ is the a~priori rotation matrix, 
$\vec{q}_e(t)$ is the empirical small vector of perturbational rotation. 
It should be noted that such a transformation does not determine coordinates
of position vectors in the terrestrial coordinate system uniquely. 
Equation \ref{e:01} should be accompanied with the vector equation 
of additional boundary conditions:
\begin{eqnarray}
  \displaystyle \sum_i^N \vec{q}_e(t) \times \vec{d}_e(t)_i = \mbox{const}
\label{e:02}\end{eqnarray}
  where summing is performed over all stations of the network. The rotation
matrix $\widehat{\mathstrut\cal M}_a(t)$ and its perturbation $\vec{q}_e(t)$
can be considered as the matrix of Earth's rotation if a) the relative motion
of stations with respect to the local areas of the Earth's crust is negligible;
b) the reference positions of observed bodies are at reset in an inertial
coordinate system. Thus, the Earth's rotation is defined here as a net 
rotation of a polyhedron of reference positions of observing stations with 
respect to a polyhedron of reference positions of observed bodies. 

  This definition of the Earth's rotation follows the materialistic 
approach: ``the Earth's rotation is that what {\bf is observed}''. 
The currently prevailing paradigm tends to consider this phenomena from 
the idealistic point of view, which involves considering a relative motion 
of celestial spheres, one of which represents the ideal Earth and another 
represents the sky (refer to \citep{r:moritz} more details). That approach 
operates such notions as big circles, poles, axes, ecliptic, true equator 
of date etc., and follows the tradition which can be traced back to ancient 
Greece \citep{r:hist}. This concept has a mechanical interpretation of 
a rolling and sliding Poinsot's cones \citep{r:poinsot}. However, this 
concept is not adequate when one needs to describe the Earth's rotation with 
an accuracy comparable to the precision of modern observations, i.e. 
$\mbox{3--5} \cdot 10^{-10}$~rad. First, the Earth can not be considered as
a rigid body, and a notion of the rotation axis, i.e. a set of points to which
a distance from any point that belongs to the body does not change with 
time, is not applicable any more. Second, the presence of high frequency
variations in the Earth's rotation, others than retrograde diurnal terms, 
makes the Poinsot's interpretation inadequate. Thirdly, the complexity of the
mathematical model --- the so-call MHB2000 expansion \citep{r:mhb2000} 
consists of more than one thousand separate motions --- makes a meaningful
interpretation problematic, similar to the situation of description of 
the planetary motion in terms of Ptolemy's epicycles.

\section{Parameterization of the Earth's rotation}

   Since both $\widehat{\mathstrut\cal M}(t)$ and vector $\vec{d}(t)$ are 
functions of time, i.e. infinite sets of points, their evaluation from
a finite set of observations can be performed only in the form of representing
them via some functions. The choice of these functions we call 
a mathematical model. The model depends on a finite set of unknown parameters 
which are determined from observations. 

   The fundamental problem is that no observation technique, except the laser
gyroscope, is sensitive to the instantaneous Earth's rotation vector or 
its time derivatives \emph{directly}. The rotation angles can be derived 
using the least square estimation procedure together with evaluation of other 
parameters, and it requires the accumulation of the sufficient amount of data 
in order to separate variables. The estimates of the Earth's rotation angles 
cannot be sampled too fast. A typical sampling rate of estimates is one day, 
since this allows compensation of a certain type of systematic errors. In some 
cases the sampling rate can be reduced to several hours.

\subsection{Traditional approach}

  A careful examination of the traditional approach first proposed 
by \citet{r:Her86}, reveals three mathematical models \citep{r:iers2003}: 
1) the a~priori model, 2) the estimation model, and 3) the post-processing 
model. Corrections to rotation angles around coordinate axes $A_i(t)$ are 
parameterized in the form equivalent to 
\begin{eqnarray}
   \begin{array}{l @{\enskip} l @{\enskip} l @{\enskip} l @{\enskip} 
                 l @{\enskip} l @{\:} l }
      A_1(t) & = &  c(t) \, \cos - \Omega_n t & + &
                    s(t) \, \sin - \Omega_n t & + \\
             &   &  b_1(t) & + & \dot{b}_1(t)*(t-t_0) &  \\
      A_2(t) & = &  c(t) \, \sin - \Omega_n t & - & 
                    s(t) \, \cos - \Omega_n t & + \\
             &   &  b_2(t) & + & \dot{b}_2(t)*(t-t_0)    \\
      A_3(t) & = &  b_3(t) & + & \dot{b}_3(t)*(t-t_0) 
   \end{array}
\label{e:03}\end{eqnarray}
  where $\Omega_n$ is the Earth's angular velocity.

  This estimation model is not adequate for a long period of time. Usually
parameters $s, c, b, \dot{b}$ are determined for a given 
24~hour period. Then the time series of $s_i, c_i, b_i$ are filtered and
smoothed. The coefficients of the interpolating spline are computed using 
the smoothed time series of these parameters. This interpolating spline
for continuous functions $s(t), c(t), b(t)$ defined at the entire interval
of observations forms the post-processing model. The a~priori rotation matrix 
$\widehat{\mathstrut\cal M}_a(t)$ is represented as a product of 12 elementary
rotations around coordinate axes. Some angles of these rotations depend on
functions $s(t), c(t), b(t)$ from  the post-processing model. 

  The traditional approach has certain disadvantages. The three mathematical
models involved contradict each other. The parameter estimation is not optimal: 
the raw time series minimizes residuals in the least square sense, but the
filtered and smoothed series does not; if it did, no smoothing would have been
needed. The estimation model \ref{e:03} has a very small range of validity.
In order to compensate its coarseness, a very refined and sophisticated
a~priori model is required. It should be accurate at a very high level.
Small changes in the a~priori model results in changes not only
adjustments, but in total Earth orientation parameters due to inadequacy of 
the estimation model \ref{e:03}. To demonstrate it, a series of VLBI solutions
was made. The first reference solution, 
gsf2006d\footnote{\tt{http://vlbi.gsfc.nasa.gov/soltions/2006d}}, used as 
a~priori the USNO Final time series of pole coordinates and UT1 with the time 
span of 1~day\footnote{\tt{ftp://maia.usno.navy.mil/ser7/finals.all}}. 
Eight parameters, $s, c, b, \dot{b}$, were estimated for each 24~hour 
observing session independently. In the second solution the Gaussian noise 
with the standard deviation 1~nrad was added to the USNO Finals EOP. 
In other trial solutions the Gaussian noise was passed through the rectangular 
low-pass digital filter with the frequency cutoffs which correspond to periods 
of 3, 7, 10 and 15 days. The noise was re-scaled in order to have the standard 
deviation of 1~nrad in all cases. The rms of the differences in \emph{totals}, 
the sum of apriori and adjustments, with respect to the reference solution 
are presented in table~\ref{t:t1}.
\begin{table}
   \caption{The rms of the differences in total EOP of solutions with 
            a~priori time series with added Gaussian noise with respect to the 
            reference solution in nrad. The Gaussian noise was passed though
            the rectangular filter which cut the frequencies that corresponding
            periods shorter than a threshold. The rms of the noise was 1~nrad
            in all cases.}
   \begin{tabular}{l @{$\:\:$} c c c}
      \hline
      Freq. cutoff       & Pole coord. &  UT1  & Nutation $ \Delta\epsilon $ \\
      \hline
      Periods $ < 1^d  $ &  0.147       & 0.138  & 0.148      \\
      Periods $ < 3^d  $ &  0.082       & 0.074  & 0.084      \\
      Periods $ < 7^d  $ &  0.048       & 0.021  & 0.016      \\
      Periods $ < 10^d $ &  0.041       & 0.021  & 0.008      \\
      Periods $ < 15^d $ &  0.038       & 0.012  & 0.004      \\
      \hline
   \end{tabular}
   \label{t:t1}
\end{table}

  This factor is one of the reasons of the so-called ``analysis noise'' --- 
the differences in results of analysis of the same data processed by 
different analysis centers. In order to reduce the influence of errors in
a~priori to a negligible level, say 1/2 of the formal uncertainty, which
is nowadays is at a level of 0.3 nrad, the errors in the a~priori EOPs 
must be very small, less than 1~nrad.

  Another disadvantage of the traditional approach is that results of 
parameter estimation are not usable \emph{directly} by an end-user: they 
should be post-processed. This makes it difficult to assess the errors 
of the interpolated time series of $c(t), s(t), b(t)$. Correlations are lost, 
contribution of errors of the a~priori EOP model is not taken into account. 
This makes evaluation of statistical hypothesis based on estimates of the 
Earth orientation parameters problematic.

\subsection{Alternative approach: expansion of the perturbational rotation
into basis functions}

  We can overcome the setbacks of the traditional approach if we refine 
the estimation model and make it valid not only over a 24~hour period, but
over the entire time range of observations, i.e. several decades. This can
be done in a form of expansion of the vector of perturbational rotation 
$\vec{q}_e(t)$ into series over basis functions. The choice of basis 
functions is determined by the nature of the process under investigation. 
It is proposed to expand $\vec{q}_e(t)$ into Fourier and B-spline bases
in this paper:
\begin{eqnarray}
   \vec{q}_e(t) = \left(
     \begin{array}{l}
       \displaystyle\sum_{k=1-m}^{n-1} f_{1k} \, B_k^m(t) \\
       + \displaystyle\sum_{j=1}^{N} \left( P^c_{j} \cos \omega_m \, t \: + \:
                                        P^s_{j} \sin \omega_j \, t \right)   
       \vspace{0.5ex} \\
         + t \, \left( S^c \cos -\Omega_n \, t \: + \:
                       S^s \sin -\Omega_n \, t \right) 
       \vspace{3ex} \\
       \displaystyle\sum_{k=1-m}^{n-1} f_{2k} \, B_k^m(t) \\
       + \displaystyle\sum_{j=1}^{N} \left( P^c_{j} \sin \omega_j \, t \: - \:
                             P^s_{j} \cos \omega_j \, t \right)
       \vspace{0.5ex} \\
         + t \, \left( S^c \sin -\Omega_n \, t \: - \:
                       S^s \cos -\Omega_n \, t \right)
       \vspace{3ex} \\
       \displaystyle\sum_{k=1-m}^{n-1} f_{3k} \, B_k^m(t) \: + \\
       \displaystyle\sum_{j=1}^{N} \left( E^c_{j} \cos \omega_j \, t + 
                             E^s_{j} \sin \omega_j \, t \right) 
       \\
     \end{array}
     \right)
\label{e:04}\end{eqnarray}
  where $B_k^m(t)$ is the B-spline function \citep{r:Nue} of degree $m$ 
determined at a sequence of knots $ t_{1-m}, \, t_{2-m}, \, \ldots, \, 
t_0, \, t_1, \, \ldots \, t_k$; $\omega_j$ are the frequencies of 
external forces; the coefficients $f_{ik}, P^c_j, P^s_j, S^c, S^s, 
E^c_{j}, E^s_{j}$ are the parameters of the expansion; $\Omega_n$ is 
the nominal frequency of the Earth's rotation 
\mbox{$ 7.292\,115\,146\,707 \cdot 10^{-5} \: \mbox{rad} \; \mbox{s}^{-1}$}.

  The B-spline basis is adequate for modeling a smooth, slow component in
$\vec{q}_e(t)$, the Fourier basis is adequate for modeling harmonic variations,
the coefficients of the cross-term, $S^c, S^s$ take into account corrections
to the precession rate and the ecliptic obliquity rate. 

  Although the B-spline and Fourier bases alone are orthogonal, the sum of
two bases is in general not orthogonal. The estimation model should be 
complemented by the orthogonality condition:
{
\small
\begin{eqnarray}
   \begin{array}{lcr}
      \displaystyle\int\limits_{t_0}^{t_1} \left( 
        \sum_{k=1-m}^{n-1} f_k \, B_k^m(t) \: \cdot \:
        \displaystyle \sum_{j}^{N} P^c_{j} \cos \omega_j \, t \right) 
        \: dt = 0 \hspace{-3em} \\
      \displaystyle\int\limits_{t_0}^{t_1} \left( 
        \sum_{k=1-m}^{n-1} f_k \, B_k^m(t) \: \cdot \:
        \displaystyle \sum_{j}^{N} P^c_{j} \sin \omega_j \, t \right) 
        \: dt = 0 \hspace{-3em} \\
   \end{array}
\label{e:05}\end{eqnarray}
}

  The choice of the degree of B-spline and the time span between knots, 
the number of constituents and the frequencies of the Fourier basis elements 
are determined by the targeted accuracy of the estimation model.
It was demonstrated by \citet{r:erm}, that the time span 3~days for $q_1(t),
q_2(t)$, 1~day for the $q_3(t)$ component, and 740~harmonic constituents 
is sufficient for keeping the accuracy of the estimation model below 
a $10^{-11}$~rad level. The frequencies of the Fourier constituents are 
selected in such a way that they correspond to excitation caused by 
a)~tide-generating potential; b)~the near diurnal free wobble; 
c)~the atmospheric nutations. 

   There are several approaches for selecting the frequencies of relevant 
Fourier constituents. In the first approach a theory of the Earth's nutations
is used. Since the spectrum of the tide generating potential consists 
of the series of sharp peaks, including constituents with the amplitudes 
of the potential and forced nutations higher than a certain limit 
is sufficient. The near diurnal free wobble and the atmospheric nutation 
are band-limited processes, so selecting all the frequencies within 
those bands with a step reciprocal to the length of observations 
is sufficient for modeling these components.

  Some components of the theoretical spectrum have a frequency separation less 
than $1/T$, where $T$ is the time interval of observations. In this case 
strong constraints are imposed to force the ratio of complex amplitudes 
to be the same as theoretical. 

  Though exploiting theoretical knowledge about the Earth rotation allows to 
reduce considerably the number of constituents of the Fourier basis, this 
raises a certain concern. If the estimation model implicitly incorporates 
theoretical assumptions, strictly speaking these estimates cannot 
be used for validation of the theory. Although the ratios of amplitudes 
between close constituents of EOP variations caused by the tidal potential 
exerted by external bodies have a sound theoretical basis, we should bear 
in mind that the ultimate goal of comparison of theoretical 
predictions with observations is to check validity of assumptions put 
in the foundation of the theory and to make a judgment whether the model is
complete or not. If there are unaccounted additive constituents at these 
frequencies, for example, caused by the free motion, by the atmospheric or 
oceanic excitation, the theoretical ratios may not be valid.

  Anther way to find the sequence of frequencies where the signal is present
is to make a set of trial solutions and to estimate amplitudes of harmonic
variations of $\vec{q}_e(t)$ at a set of frequencies sampled with a step 
reciprocal to the length of observations within the diurnal, semi-diurnal 
and ter-diurnal bands and discard the constituents where no statistically
significant signal was detected. This approach is free from a potential
bias of adopted theories.

\section{Estimation of the vector of the perturbational rotation from VLBI
         data}

  The VLBI dataset from January 1984 through September 2006 was used for 
validation of the proposed approach. On average, ~150 twenty four experiments
per year have been observed. The number of participating stations in each 
individual session varies from 2 to 20, although 4--7 is a typical number. 
No station participated in all sessions, but every station participated 
in sessions with many different networks. All networks have common nodes and, 
therefore, are tied together. 


  The requirement to the accuracy of the a~priori Earth's rotation model 
is determined by a condition that $||\vec{q}_e||^2$ be less than errors
of determination of the perturbational vector $\vec{q}_e$, i.e. 
\mbox{$10^{-11}$ rad}. This gives the requirement of the accuracy of the 
a~priori matrix: \mbox{$3 \cdot 10^{-6}$~rad}. This is three orders
of magnitude lower than the requirement to accuracy of the a~priori 
rotation matrix in accordance with the traditional approach. We can exploit 
this advantage of the alternative approach and use the simplest possible 
model. The following expression for the a~priori matrix of the Earth's 
rotation $ \widehat{\mathstrut\cal M}_a(t) $ according to the Newcomb-Andoyer 
formalism was used:
\begin{eqnarray}
   \begin{array}{@{\!} l@{\,} l}
      \widehat{\mathstrut\cal M}_a(t) = & 
      \widehat{\mathstrut\cal R}_3(\zeta_0)  \cdot
      \widehat{\mathstrut\cal R}_2(\theta_0)  \cdot
      \widehat{\mathstrut\cal R}_3(z)  \cdot
      \widehat{\mathstrut\cal R}_1(-\epsilon_0)    \cdot   
      \hphantom{-}\hspace{-3em}\hphantom{-} \\ &
      \widehat{\mathstrut\cal R}_3(\Delta\psi)  \, \cdot 
      \widehat{\mathstrut\cal R}_1(\epsilon_0 + \Delta\epsilon) \cdot 
      \widehat{\mathstrut\cal R}_3(-S) 
      \hphantom{-}\hspace{-3em}\hphantom{-}
   \end{array}
\label{e:06}\end{eqnarray}
  where $ \widehat{\mathstrut\cal R}_i $ is a rotation matrix around
the axis $i$. For the variables $\zeta_0, \theta_0, z, \epsilon_0, \Delta\psi,
\Delta\epsilon_0, S, $ the following simplified expression were used:

\begin{eqnarray}
   \begin{array}{lcl}
      \zeta_0    & = & \zeta_{00} \: + \: \zeta_{01} \, t \: + \:
                       \zeta_{02} \; t^2 \\
      \theta_0   & = & \theta_{00} \: + \: \theta_{01} \, t \: + \: 
                       \theta_{02} \, t^2 \\
       z         & = & z_{0\hphantom{0}} \: + \: z_{1\hphantom{0}} \, t \: + \: 
                       z_{2\hphantom{0}} \, t^2  \\
      \epsilon_0 & = & \epsilon_{00} \: + \: \epsilon_{01} \, t \: + \:
                       \epsilon_{02} \, t^2                         \\
      \Delta\psi & = & \displaystyle \sum_j^2
                       p_j \, \sin \ ( \alpha_{j} + \beta_{j} \, t ) \\
                       \hphantom{-}\hspace{-5em}\hphantom{-} \vspace{-3ex} \\
   \end{array}
\label{e:07}\end{eqnarray}

\begin{eqnarray}
\begin{array}{lcl}
      \Delta\epsilon
                 & = & \displaystyle \sum_j^2 
                        e_j \, \cos \ ( \alpha_{j} + \beta_j \, t ) 
                       \hphantom{-}\hspace{-5em}\hphantom{-} \\
      S   & = & S_0 + E_0 \: + \;  \Delta\psi \, \cos\epsilon_0 \\ & & + \: 
                (\Omega_n + \zeta_{01} + z_1 + E_1 ) \, t \\ & & + \: 
                    (\zeta_{02} + z_2 + E_2 ) \, t^2
                       \hphantom{-}\hspace{-5em}\hphantom{-} \\
          &   & + \:
                    \displaystyle
                    \sum_i^2 \left( E^c_i \cos \gamma_{i} \,t + 
                                    E^s_i \sin \gamma_{i} \, t \right) 
   \end{array}
\nonumber\end{eqnarray}
%
  Here $t$ is TAI time elapsed from 2000, January 01, 12 hours. Some of these 
parameters were taken from theory, some of them were found with LSQ fit of 
time series of adjustments of pole coordinates and UT1. Errors of this apriori 
Earth's rotation model are less than $2 \cdot 10^{-6}$ rad over the period 
1984--2006. 

  Unlike to processing GPS observations, analysis of VLBI observations is 
done in a so-called global mode: a set of 2,000--20,000 global parameters 
which are considered common over the entire set of observations, i.e. 
22~years; 1,000--50,000 local parameters specific for a given 24 hour 
session; and over a million of segmented nuisance parameters specific 
for a 20--60 minute interval are solved in a single least square solution 
directly. 

  In the present solution global parameters were stations positions, station 
velocities, amplitudes of harmonic variation in site positions, coefficients 
of B-spline modeling non-linear site position variations, source coordinates 
and proper motions, harmonic variations at 740 frequencies and coefficients 
of B-spline of the perturbational vector of the Earth's rotation $\vec{q}_e$.

\subsection{Summary of the VLBI results}

   Results of analysis are available on the Web at 
{\tt http://vlbi.gsfc.nasa.gov/erm}. The weighted root mean square of 
postfit residuals was the same as in the solution that followed the 
traditional approach, 21.9~ps. 

   For comparison with the USNO time series of pole coordinates and UT1-TAI,
these time series were transformed to the vectors of perturbational rotation, 
and the coefficients of the cubic interpolating spline was computed. These
coefficients represents the USNO mathematical model of the Earth rotation.
The accuracy of the empirical Earth rotation model is higher at the instants 
of time of middle of observing sessions. The values of vectors of perturbational 
rotation and their time derivatives were computed at these moments of time 
for both the empirical model from the VLBI solution and from the USNO 
mathematical model. The rms of these differences are presented in 
table~\ref{t:t2}. It should be noted that the GPS estimates of pole coordinate
almost entirely dominate the USNO time series. Therefore, the differences
between pole coordinates and their rates give us a measure of an agreement 
between the empirical Earth rotation model from VLBI and the \emph{independent} 
estimates from GPS.
\begin{table}
   \caption{The first row shows the rms of the differences over the period
            1996.0--2006.0 between two models of the Earth's rotation: the 
            empirical model and the USNO model. The second row shows the 
            rms of the differences between solution gsf2006d that follows
            the traditional approach and the USNO model.}
   \begin{tabular}{l c c c c}
      \hline
      Solution & Pole & UT1  & Pole rate & UT1 rate \\
               & \multicolumn{2}{c}{$10^{-9}$ rad } 
               & \multicolumn{2}{c}{$10^{-14}$ rad s${}^{-1}$} \\
      \hline
       ERM      & $ 0.64 $  &  $ 0.52 $ 
                & $ 0.99 $  &  $ 0.81 $ \\
       gsf2006d & $ 0.56 $  &  $ 0.42 $ 
                & $ 1.95 $  &  $ 1.52 $ \\
      \hline
   \end{tabular}
   \label{t:t2}
\end{table}

  In order to validate the estimates of the harmonic variations of the 
perturbational rotation vector, a trial solution following the traditional
Earth rotation parameterization and the a~priori empirical model of the
Earth rotation from the previous solution was run. The rms of adjustments 
of nutation angles over 1996.0--2006.0 with respect to the apriori MHB2000 
expansion~\citep{r:mhb2000} was 0.98~nrad and 0.39~nrad with respect 
to the empirical Earth rotation model.

\section{Conclusions}

  It was demonstrated that the empirical Earth rotation model can be determined 
directly from observations over a period of 22 years using the least square 
estimation technique. The advantage of the proposed approach is that 
a continuous function describing the Earth's orientation is determined in one 
step without producing intermediate time series. Another advantage of the 
proposed approach is that a simplified a~priori model with only 31 numerical 
parameters is sufficient, while according to the traditional approach 
a complicated a~priori model of precession, nutation, high frequency harmonic 
variations of the Earth's rotation and a filtered and smoothed time series 
of the Earth orientation parameters produced in the previous analysis, 
in total $46\,000$ numerical parameters \citep{r:iers2003}, is needed. 

  The proposed approach is conceptually much more simple than the traditional
approach, since it does not operate with idealistic notions that are not 
observable, such as the non-rotating origin, the equinox, various intermediate 
poles, axes, etc.

  The EOP, the station position and velocities, the source coordinates are
produced in a single LSQ adjustment, and therefore, are mutually consistent.

  It was demonstrated that the empirical Earth rotation model derived from 
analysis of VLBI observations gives the differences with respect to the EOP 
derived from analysis of independent GPS observations at moments of observation
at the same level, within 20\%, as the differences of the VLBI EOP series 
produced with the traditional approach. The advantage of the proposed approach
is that the estimates of the EOP rates are by a factor of 1.5--2.0 closer to
the GPS time series than the VLBI EOP rates estimated following the traditional 
approach. The rms of estimates of residual nutation angles with respect to the 
empirical Earth rotation model is 2.6 times smaller than the residuals with
respect to the MHB2000 expansion.

\end{document}